\documentclass[aps,pra,onecolumn,groupedaddress,showpacs]{revtex4}

\usepackage{amssymb,amsmath}
\usepackage{graphicx}

\newcommand{\Om}{\Omega}
\newcommand{\om}{\omega}
\newcommand{\ve}{\varepsilon}
\newcommand{\pa}{\partial}
\DeclareMathOperator{\sech}{sech}

\begin{document}

\title{Light-Matter Interaction and Hybrid Vector Breather}

\author{G. T. Adamashvili}
\affiliation{Technical University of Georgia, Kostava str.77, Tbilisi, 0179, Georgia.\\ email: $guram_{-}adamashvili@ymail.com.$ }

\begin{abstract}
The nonlinear coherent interaction of light with the dispersive and Kerr-type third-order susceptibility medium containing optical impurity atoms or semiconductor quantum dots is considered. Using the generalized perturbation reduction method, the nonlinear wave equation is reduced to the coupled nonlinear Schr\"odinger equations. It is shown that the second-order derivatives play a key role  in the description of the process of formation of the bound state of two breathers oscillating with the sum and the difference of frequencies and wave numbers.
The resonant, nonresonant and hybrid  mechanisms of the  formation of the two-component nonlinear pulse - the vector  breather are realized depending on the light and medium parameters.
 Explicit analytical expressions for the profile and parameters of the nonlinear pulse  are presented. The conditions of the excitation of resonant, nonresonant and hybrid nonlinear waves are discussed. In the particular case, the resonant vector breather coincides with the vector $0\pi$ pulse of self-induced transparency.
\end{abstract}


\maketitle

\section{Introduction}

The nonlinear light-matter interaction is a field of intensive research that covers a wide class of nonlinear optical phenomena.  At the nonlinear coherent interaction of light and matter there occur various nonlinear processes. A special interest is shown in the processes which lead to the formation of nonlinear solitary waves of stable profile. The propagation of these waves is one of the most striking demonstrations of nonlinearity in optical media. The analysis of the mechanisms causing the formation of nonlinear solitary optical waves, the establishment of their types and the study of their properties in various nonlinear media are among the basic problems of the physics of nonlinear optical waves.

Depending on the character of nonlinearity, the nonresonant and resonant mechanisms of the formation of nonlinear solitary waves are considered. In the case of nonresonant nonlinearity, which can be connected with the quadratic (second-order) or cubic (third-order) nonlinear susceptibilities, its competition with dispersion or diffraction leads to the formation of nonresonant optical nonlinear waves of stable profiles which can be respectively classified as temporal  or spatial solitary waves \cite{Sauter::96,Suchorukov::88,Vinogradova::90,Akhmediev::1997, KivAg::03, AdamashviliMaradudin:PhysRevE:97,LambJr::80, Maimis::99}. Resonant optical solitary waves can be formed in the conditions of self-induced transparency (SIT) \cite{McCall:PhysRev:69,Maimistov:PhysRep:90,Allen::75, Dodd::1982,Diels:PhysRev:74, Poluektov:UspFizNauk:74, LambJr:RevModPhys:71, Crisp:PhysRep:70, Rothenberg::1984,Harvey::94, Giessen:PhysRevLett:98}.

Both  resonant  and nonresonant nonlinear  solitary  waves are  divided  into  two  basic  types:  single-component  and two-component nonlinear  waves.  In the  main,  to single-component nonlinear  solitary  waves  are attributed  scalar  solitons  and  scalar breathers, and to two-component waves there belong vector  solitons and  vector  breathers or their  modifications.

Although  the  SIT phenomenon and resonant nonlinear  waves, namely,   scalar  solitons and scalar  breathers have already been  studied for quite a  long time,  in recent years  the interest in their  investigation has noticeably revived. A lot of theoretical and experimental studies have been carried out \cite{Schneider:ApplPhysLett:03, Adamashvili:PhysLettA:07, Adamashvili:PhysRevA:2007, AKK:Phys.Rev.A:08, Panzarini:PhysRevB:02, Adamashvili:Eur.Phys.J.D.:08, Adamashvili:PhysRevA:14, Adamashvili:TPL:09, Adamashvili:OS:09, Arkhipov::2016, Adamashvili:OptLett:06, Arkhipove::2020, Arkhipove::2019, Arkhipove::19, Zabolotskii::03, Sazonov::06, Sazonov::18, Ustinov::06, Steudel::05, Rozanov::10, Visotina:Pisma v JTEP:2006, Arkhipov::2018, Visotina:Optics and spectroscopy:2006, Akhmediev:Diss:2005,Ivic::2019}.

In  SIT theory,  second-order  derivatives in  space coordinate and the time   of the electric field strength of the pulse in the Maxwell wave equation  were usually   neglected. So, it was believed that the basic SIT waves are the scalar single-component $0\pi$ pulse and the scalar singe-component McCall-Hahn's $2\pi$ pulse \cite{McCall:PhysRev:69, Maimistov:PhysRep:90, Allen::75, Poluektov:UspFizNauk:74, Dodd::1982,Diels:PhysRev:74, LambJr:RevModPhys:71}. However, the situation has recently changed due to the development of a new mathematical approach - the generalized perturbative reduction method (PRM) \cite{Adamashvili:Result:11, Adamashvili:Optics and spectroscopy:2012, Adamashvili:Physica B:12, Adamashvili:Eur.Phys.J.D.:12, Adamshvili:Arxiv:2014, Adamashvili:PhysLettA:2015, Adamashvili:Optics and spectroscopy:2019, Adamashvili:Eur.Phys.J.D.:20, Adamashvili:PRE:16, Adamashvili:PRE:12,  Adamshvili:Arxiv:2019, Adamshvili:Arxiv:2020}.

As different from the standard PRM  \cite{Taniuti::1973} adapted for the consideration of  single-component nonlinear  waves,  the  generalized  PRM  makes it possible to  proceed  to  the  next stage  of  the  development  of  SIT  theory,  i.e.  to the study of the  properties   of two-component  solitary waves. The generalized PRM demonstrates what an important role the second-order derivatives of the Maxwell wave equation play in SIT theory.  These derivatives are used, in particular, to describe the interaction between two single-component breathers, which leads to the formation of their bound state.  Using the generalized PRM, qualitatively new results were obtained in SIT theory.  It was established that there exists a two-component vector breather (TVB) which in SIT theory is called the vector $0\pi$  pulse. The  vector  $0\pi$ pulse is the  bound  state  of  two breathers of the  same polarization, one component oscillating  with  the sum and the other with the difference of  frequencies and wave numbers (SDFW). As a result of such a superposition there arises a   nonlinear  zero-area pulse with  specific phase modulation and the profile which significantly  differs from the profiles of the scalar single-component SIT-soliton  and the scalar single-component SIT-breather. This clearly implies that the basic SIT pulse is the vector $0\pi$  pulse and the scalar $2\pi$ pulse.  But the scalar $0\pi$  pulse of SIT is only a certain  approximation of the vector  $0\pi$  pulse, which takes  place only if the  second-order derivatives in SIT equations  are neglected or a less general  mathematical approach than  the generalized  PRM,  for instance,  the standard PRM  is used (for details see \cite{Adamashvili:Optics and spectroscopy:2019,Adamashvili:Eur.Phys.J.D.:20, Adamshvili:Arxiv:2019}).

The optical resonant vector $0\pi$ pulse in various isotropic and anisotropic materials was investigated for one-photon and two-photon resonance excitations, plane and surface waves, also the waveguide modes were studied in different physical situations \cite{Adamashvili:Result:11, Adamashvili:Optics and spectroscopy:2012, Adamashvili:Physica B:12,Adamshvili:Arxiv:2014, Adamashvili:Eur.Phys.J.D.:12, Adamashvili:PhysLettA:2015}. Optical nonresonant TVB oscillating with the SDFW in the dispersive and Kerr-type third-order susceptibility medium were considered using the generalized PRM \cite{Adamshvili:Arxiv:2018}.

Depending  on  the  numerical  values  of light and medium  parameters there may occur  physical  situations in which  both  resonant and nonresonant mechanisms  of the formation  of nonlinear  waves act   simultaneously. In that case, the  hybrid  (blended) mechanism  of excitation of a  nonlinear  solitary  wave becomes active and a hybrid  nonlinear  pulse  can be formed.   Hybrid  single-component scalar  solitons  and  hybrid  single-component scalar breathers were investigated by different mathematical approaches  in the course of many years \cite{Maimistov::83,Caetano::03,Nakazawa::91,Nakazawa2::91,Adamashvili:PLA:03,Fonseca::01,Adamashvili:PhysRevE:04,Adamashvili:PRE:04,Adamashvili:PhysRevE:06,Adamashvili:Phys.Rev.A:19}.
 However, using the generalized  PRM  we have obtained  the novel result - the hybrid TVB oscillating with the SDFW which has not been studied until now.

The  goals of the present study are as follows:  investigation of processes of the formation  of the optical  hybrid  TVB  oscillating  with  the SDFW and propagating in dispersive Kerr-type nonlinear medium  containing a small concentration of optical impurity atoms  or semiconductor quantum dots  (SQDs); derivation of explicit  analytical  expressions  for the parameters of the hybrid TVB oscillating with the  SDFW in the carrier wave frequency and wave number region; definition of the conditions for the existence  of the nonlinear  waves.

The rest of this paper is organized as follows.   Section  II  is devoted  to  the  derivation of SIT  equations  in dispersive Kerr-type medium  containing  optically  active  impurity atoms  (SQDs)  for slowly varying  envelope functions.   In Section  III, using the  generalized  PRM  the  nonlinear  wave equation for these functions is transformed to coupled  nonlinear Schr\"odinger equations (NSEs)  for  auxiliary  functions.   The explicit expressions are obtained for the TVB oscillating with the SDFW. In the  last  section  IV, we discuss the obtained results, the  role of  second-order  derivatives  in the  process of  formation of the  bound  state  of two wave packets  and  consider the criteria  of the existence  of  resonant,  nonresonant and hybrid  nonlinear  waves.

\vskip+0.5cm

\section{Basic equations}

We study the physical process of formation of the optical TVB oscillating with the SDFW in the dispersive  Kerr-type third-order susceptibility medium containing a small concentration $n_0$ of optical active impurity atoms or SQDs. The optical linear-polarized along the $\emph{x}$ axis pulse with width $T<<T_{1,2}$, frequency $\omega >>T^{-1}$ and wave vector $\vec{k}$  spreads in the positive direction along the $\emph{z}$-axis. Here $\omega$ is the carrier wave frequency of the optical wave, while  $T_1$ and $T_2$ are  respectively the longitudinal and the transverse relaxation time of optical resonant atoms or SQDs. In general, the dispersion can be described by means of the electric permittivity tensor $\varepsilon_{ij}(z,t)$ which depends on the space coordinate $\emph{z}$ and the time $\emph{t}$. But  in  an isotropic medium the permittivity tensor  reduces to the form $\varepsilon_{ij}(z,t)=\varepsilon(z,t)\delta_{ij}$, where $\varepsilon(z,t)$ is the scalar function, $\delta_{ij}$ is the Kronecker symbol.

The wave equation for the $\emph{x}$ -component of the strength of the electric field $\vec{E}(E,0,0)$  is written in the form \cite{Landau:Electrodynamics :84, Vinogradova::90}
\begin{equation}\label{eq}
 \mathfrak{c}^2 \frac{\pa^{2} E}{\pa {z}^2} -\frac{\pa^{2} D_{l}}{\pa t^2}=4\pi\frac{\pa^{2} P}{\pa t^2},
\end{equation}
where $\mathfrak{c}$ is the velocity of light in vacuum, $D_{l}$ is the linear part of the $\emph{x}$ -component of the electric displacement vector (Appendix I). The nonlinear polarization of the medium $P=P_{n}+P_{r},$ contains the nonresonant and resonant parts. Here
\begin{equation}\label{2}
P_{n}=\int\rho_{xxxx}({z}_1,{z}_2,{z}_3,t_1,t_2,t_3)E(z-z_1,t-t_1)
E(z-z_1-z_2,t-t_1-t_2)\times$$$$ E(z-z_1-z_2-z_3,t-t_1-t_2-t_3)  dz_1 dz_2
dz_3 dt_1 dt_2 dt_3
\end{equation}
is the $\emph{x}$ -component of the nonresonant nonlinear Kerr-type  polarization  of the medium, $\rho_{xxxx}$ is the component of the third-order susceptibility tensor.
$P_{r}=n_0 \mu s_1$ is the $\emph{x}$ -component of the polarization of two-level optical impurity atoms (SQDs), where $\mu $ is the dipole matrix element of the optical atoms (SQDs), $s_i(t)=<\hat{\sigma}_i(t)>$ are average quantities of the Pauly's operators $\hat{\sigma}_i,$ $\;(i=1,2,3)$.

The dependence of the function $\emph{P}_{r}$  on the strength of the electrical field $\emph{E}$ is defined by the optical Bloch equations \cite{Allen::75, Landau:Quantum Mechanics:80, Adamashvili:PhysRevA:2007}
$$
    \frac{\pa s_1}{\pa t}=-\om_0 s_2,
$$
$$
\frac{\pa s_2}{\pa t}=\om_0 s_1+ {\kappa}_0 E s_3,
$$
\begin{equation}\label{bl}
    \frac{\pa s_3}{\pa t}=-{\kappa}_0 Es_2,
\end{equation}
where
${\kappa}_0=\frac{2\mu}{\hbar},$ $\hbar$ is Planck's constant, $\om_0$ is  the  excitation frequency  of optical impurity two-level atoms.

The system of equations  (1)-(3), which are SIT equations in the dispersive Kerr-type medium, can be simplified by using the method of
slowly varying profile. To this end, we write the $\emph{x}$ -component of the strength of the electric field $\emph{E}$  and the polarization $\emph{P}_{r}$ in the form \cite{Maimistov:PhysRep:90, Allen::75, Poluektov:UspFizNauk:74, Diels:PhysRev:74, Dodd::1982,LambJr:RevModPhys:71}
\begin{equation}\label{E}
 E=\sum_{l=\pm1}\hat{E}_{l} Z_l,
\;\;\;\;\;\;\;\;\;\;\;\;\;\;\;\;
P_{r}=n_0 \mu \sum_{l=\pm1}  d_{-l}Z_l,
\end{equation}
where $\hat{E}_{l}$ and  $d_{l}$   are the slowly varying complex amplitudes of the optical electric field and the polarization of optical active atoms. These are complex functions in view of the fact that the wave is phase modulated.  $Z_{l}= e^{il(k z -\omega t)}$ is the fast oscillating function. Because $\emph{E}$ is a real function, we set $\hat{E}_{l}=\hat{E}_{-l}^{\ast}$.

As compared with the carrier wave parts, the envelopes $\hat{E}_{l}$ and  $d_{l}$  vary with a sufficient slowness in space and time. Therefore the following inequalities are valid
for $\hat{E}_{l}$
\begin{equation}\label{Ap}
 \left|\frac{\partial \hat{E}_{l}}{\partial t}\right|\ll\omega
|\hat{E}_{l}|,\;\;\;\left|\frac{\partial \hat{E}_{l}}{\partial z
}\right|\ll k|\hat{E}_{l}|,
\end{equation}
and similar inequalities hold for the complex function $d_{l}$.

Substituting  Eq.\eqref{E}  into the wave equation \eqref{eq}, and taking into account Eq.\eqref{D2}  (see Appendix I), we obtain  the dispersion law for the propagating  pulse in the medium
\begin{equation}\label{dis}
\mathfrak{c}^{2}k^{2}= {\om}^2{\kappa}
\end{equation}
and the nonlinear wave equation for the envelope function $\hat{E}_{l}$   in the form:
\begin{equation}\label{equa}
\sum_{l=\pm1}Z_{l}\{[ig_{1} \frac{\partial \hat{E}_{l}}{\partial z}+ig_{3} \frac{{\pa}{\hat{E}_l}}{\pa t} +g_{2} \frac{\partial^{2}\hat{E}_{l}}{\partial z^2} -g_{5} \frac{{{\pa}^2}{\hat{E}_l}}{\pa t^2} -g_{4}\frac{{{\pa}^2}{\hat{E}_l}}{{\pa {z}}{\pa  t}}]
   -4\pi \omega^{2}  \sum_{l'}\sum_{l''}  \tilde{\rho}_{l,l',l''} {\hat{E}_{l-l'-l''}}{\hat{E}_{l'}} {\hat{E}_{l''}}\} + 4\pi\frac{\pa^{2} P_{r}}{\pa t^2}=0,
\end{equation}
where
\begin{equation}\label{coef1}\nonumber\\
g_{1}={\om}^2 a-2lk \mathfrak{c}^{2},\;\;\;\;\;\;\;\;\;\;\;\;\;\;\;\;\;\;\;\;
g_{2}= {\om}^2 c -\mathfrak{c}^{2},\;\;\;\;\;\;\;\;\;\;\;\;\;\;\;\;\;\;\;\;
g_{3}=-({\om}^2 b +2l{\om}\kappa),
$$$$
g_{4}=2l{\om}a + {\om}^2 \tilde{t},\;\;\;\;\;\;\;\;\;\;\;\;\;\;\;\;\;\;\;\;g_{5}=-( {\om}^2 d+2l{\om}b+ \kappa).
$$$$
\tilde{\rho}_{l,l',l''}=\int\rho_{xxxx}({z}_1,{z}_2,{z}_3,t_1,t_2,t_3)
 e^{-il(kz_1- \om t_1)}
e^{-i(l'+l'')[k z_2- \om t_2]} e^{-il''[k z_3- \om t_3]}dz_1 dz_2
dz_3 dt_1 dt_2 dt_3.
\end{equation}

Eq.\eqref{equa} describes various nonlinear processes which arise for the nonlinear coherent interaction of light and the dispersive Kerr-type medium containing  optical impurity atoms (SQDs), and in particular,  processes of the formation of single-component and two-component solitary nonlinear waves when resonant and nonresonant mechanisms  act simultaneously.

\vskip+0.5cm

\section{Application of the generalized perturbation reduction method}

For the analyze of the two-component nonlinear solitary wave solution of Eq.\eqref{equa} we apply the generalized PRM \cite{Adamashvili:Result:11, Adamashvili:Optics and spectroscopy:2012, Adamashvili:Physica B:12, Adamashvili:Eur.Phys.J.D.:12, Adamshvili:Arxiv:2014, Adamashvili:PhysLettA:2015, Adamashvili:Optics and spectroscopy:2019} by means of which this equation can be transformed to the  coupled NSEs. In this approach the envelope function $\hat{E}_{l}(z,t)$ can be represented as:
\begin{equation}\label{ser}
\hat{E}_{l}(z,t)= \sum_{\alpha=1} \ve^\alpha
{{\hat{E}}_{l}}^{(\alpha)}=\sum_{\alpha=1}^{\infty}\sum_{n=-\infty}^{+\infty}\varepsilon^\alpha
Y_{l,n} f_{l,n}^ {(\alpha)}(\zeta,\tau),
\end{equation}
where
$$
Y_{l,n}=e^{in(Q_{l,n}z-\Omega_{l,n}
t)},\;\;\;\zeta_{l,n}=\varepsilon Q_{l,n}(z-{v_{g;}}_{l,n}
t),\;\;\;\tau=\varepsilon^2 t,\;\;\;
{v_{g;}}_{l,n}=\frac{d\Omega_{l,n}}{dQ_{l,n}},
$$
$\varepsilon$ is a small parameter. Such an expansion allows us to separate from $\hat{E}_{l}$ the even more slowly changing auxiliary function $ f_{l,n}^{(\alpha )}$. It is assumed that the quantities $\Omega_{l,n}$, $Q_{l,n}$, and $f_{l,n}^{(\alpha)}$ satisfy the inequalities:
\begin{equation}\label{ryp}\nonumber\\
\omega\gg \Omega_{l,n},\;\;k\gg Q_{l,n},\;\;\;
\end{equation}
$$
\left|\frac{\partial
f_{l,n}^{(\alpha )}}{
\partial t}\right|\ll \Omega_{l,n} \left|f_{l,n}^{(\alpha )}\right|,\;\;\left|\frac{\partial
f_{l,n}^{(\alpha )}}{\partial z }\right|\ll Q_{l,n}\left|f_{l,n}^{(\alpha )}\right|.
$$
for any value of indexes $l$ and $n$.

Although the quantities  $Q_{l,n}$, $\Omega_{l,n}$, $\zeta_{l,n}$ and ${v_{g;}}_{l,n}$ depend on  $l$ and $n$, for simplicity, we omit these indexes in the equations below when this does not cause confusion.

The generalized PRM \eqref{ser} holds for the  phase modulated complex function $\hat{E}_{l}$. Otherwise, if the wave is not phase-modulated, then the function $\hat{E}_{l}=\hat{E}_{-l}=\hat{E}$ is real and does not depend on the index $l$.

It should be noted   that the  generalized  PRM  is a sufficiently  general  mathematical approach and  can  be used  not  only  for studying two-component solitary  optical  waves, but  also  for the  investigation of two-component waves in nonlinear  acoustics,  hydro-dynamics,  plasma  physics, and so on \cite{Adamashvili:PRE:12, Adamashvili:PRE:16, Adamshvili:Arxiv:2020, Adamshvili:Arxiv:2016, Adamashvili:TPL:17, Adamashvili:AcousPhy:17}.

Substituting Eq.\eqref{ser} into  Eq.\eqref{equa}, and taking into account the explicit form of the polarization  envelope \eqref{p}, we obtain
\begin{equation}\label{9}
\sum_{l=\pm1}\sum_{\alpha=1}^{\infty}\sum_{n=-\infty}^{+\infty}\varepsilon^\alpha
Y_{l,n} Z_l\{\tilde{W}_{l,n}+ i \varepsilon J_{l,n} \frac{\partial }{\partial \zeta}
+ i\varepsilon^2 h_{l,n} \frac{\partial }{\partial \tau} +\varepsilon^{2} H_{l,n} \frac{\partial^2 }{\partial\zeta^2}\}f_{l,n}^{(\alpha)}
$$$$
=\ve^3 \sum_{{l=\pm 1}}Z_l  R_{l} [( | f_{l,l}^ {(1)}|^{2}   +  2  | f_{l,-l}^ {(1)}|^{2}  ) f_{l,l}^ {(1)}  Y_{l,l}
+(  |f_{l,-l}^ {(1)}|^{2}  + 2  | f_{l,l}^ {(1)}|^{2} )f_{l,-l}^ {(1)} Y_{l,-l}]
$$$$
 - \ve^{3}i \ae_{2} \sum_{l=\pm 1}l Z_l \int\frac{\pa {{\Theta}_{l}}^{(1)}}{\pa t}{{\Theta}_{-l}}^{(1)} {{\Theta}_{l}}^{(1)}dt'  +...\}
\end{equation}
where
\begin{equation}\label{10}
\tilde{W}_{l,n}= g_{3}n\Omega - g_{1} nQ - g_{2} Q^{2} +g_{5} {\Omega}^{2} -g_{4} Q \Omega
+\ae_{1}\frac{l n}{\Omega},
$$$$
J_{l,n} =g_{3}  v_g - g_{1} - 2 g_{2} nQ + 2 n g_{5}  \Omega  v_g
-g_{4}n(Q v_g +\Omega),
$$$$
h_{l,n}=g_{3} + 2 g_{5}n\Omega -g_{4} nQ,
$$$$
H_{l,n}=Q^{2}(g_{2} -g_{5} v_g^{2}+g_{4} v_g ),
$$$$
\ae_{1}=\frac{4\pi \omega^{2} n_0 \mu^{2}} {\hbar} <g>,
$$$$
\ae_{2}= \frac{8 \pi n_0  \mu^{4}}{\hbar^{3}} <g>,
$$$$
R_{l}=4\pi \omega^{2} (\tilde{\rho}_{l,l,l}
+\tilde{\rho}_{l,l,-l} +\tilde{\rho}_{l,-l,l}).
\end{equation}

To define the function $f_{l,n}^{(\alpha)}$ in Eq.(9), we equate to zero the terms corresponding to the same powers of $\varepsilon$. As a
result, we obtain a sequence of equations. Starting with first order of $\ve$, we find that only the components  $f _{\pm 1,\pm 1}^{(1)}$ or $f _{\pm 1,\mp 1}^{(1)}$
of the function $f_{l,n}^{(1)}$ are different from zero. The relations between the parameters $\Omega_{l,n} $ and $Q_{l,n}$ is defined by Eq.(9) and has the form
\begin{equation}\label{diss}
g_{3}n\Omega_{l,n}  -g_{1} n Q_{l,n} +\ae_{1}\frac{l n}{\Omega_{l,n} }= g_{2} Q_{l,n}^{2} -g_{5} {\Omega^{2}_{l,n} } +g_{4}Q_{l,n}  \Omega_{l,n}.
\end{equation}

Substituting  Eq.\eqref{diss} into Eq.(9), we prove that the following equations hold $J_{\pm 1,\pm 1}=J_{\pm 1,\mp 1}=0$. From Eq.(9), to third order in $\varepsilon $, and  introducing the functions   $u_{\pm 1}=\varepsilon f_{+1,\pm 1}$, we get the coupled NSEs in the form
\begin{equation}\label{cnse}
[i  (\frac{\partial u_{\pm 1}}{\partial t}+v_{\pm } \frac{\partial u_{\pm 1}}{\partial z})+  p_{\pm} \frac{\partial^{2}u_{\pm 1} }{\partial z^{2}}+ \mathfrak{q}_{\pm} |u_{\pm 1}|^{2}u_{\pm 1}  +r_{\pm}|u_{\mp 1}|^{2} u_{\pm 1}=0,
\end{equation}
where
\begin{equation}\label{13}
v_{\pm}=v_{{g;}_{+1,\pm 1}},\;\;\;\;\;\;\;\;\;\;\;
p_{\pm}= \frac{H_{+1,\pm 1}}{h_{+1,\pm 1} Q^{2}_{\pm}},\;\;\;\;\;\;\;\;\;\;\;\;\;
\mathfrak{q}_{\pm}= \frac{\kappa_{1}^{\pm}}{h_{+1,\pm 1}},\;\;\;\;\;\;\;\;\;\;
r_{\pm}= \frac{\kappa_{2}^{\pm}}{h_{+1,\pm 1}}
$$
$$
\kappa_{1}^{\pm}= R_{+1}  \pm   \frac{\ae_{2}}{\Omega_{\pm }^{3}}            ,\;\;\;\;\;\;\;\;\;\;\;\;\;\;\;\;\;\;\;\;\;\;
\kappa_{2}^{\pm}=  2 R_{+1} \pm \frac{\ae_{2}}{\Omega_{\pm }^{2}\Omega_{\mp}} ( \frac{\Omega_{\pm }}{\Omega_{\mp}}
-1 ).
\end{equation}

From Eq.\eqref{diss} it follows that
\begin{equation}\label{14}
\Omega_{+1,-1}=\Omega_{-1,+1}=\Omega_{-},\;\;\;\;\;\;\Omega_{+1,+1}=\Omega_{-1,-1}=\Omega_{+},
$$
$$
Q_{+1,-1}=Q_{-1,+1}=Q_{-},\;\;\;\;\;\;Q_{+1,+1}=Q_{-1,-1}=Q_{+}.
\end{equation}

Substituting the components of the vector soliton $u_{+1}$ and $u_{-1}$ from Eq.\eqref{uet}  into Eqs.\eqref{E} and \eqref{ser}, for the $x$-component of the electric field $E(z,t)$  we obtain the TVB oscillating with the SDFW in the form
\begin{equation}\label{vp}
E(z,t)=\frac{2}{\mathfrak{b} T}\sech(\frac{t-\frac{z}{V}}{T})\{  f_{+1} \cos[(k+Q_{+}+k_{+1})z -(\om +\Omega_{+}+\omega_{+1}) t]
+f_{-1}\cos[(k-Q_{-}+k_{-1})z -(\om -\Omega_{-}+\omega_{-1})t]\}.
\end{equation}

\vskip+0.5cm
\section{ Conclusions and discussions}

We consider the coherent nonlinear interaction of light  with the dispersive  Kerr-type medium containing optical impurity atoms or SQDs.
It is shown that along with  one-component nonlinear waves (solitons and breathers) investigated earlier \cite{Maimistov::83,Caetano::03,Nakazawa::91,Nakazawa2::91,Adamashvili:PLA:03,Fonseca::01,Adamashvili:PhysRevE:04,Adamashvili:PRE:04,Adamashvili:PhysRevE:06,Adamashvili:Phys.Rev.A:19}, the two-component nonlinear wave (vector breather) Eq.\eqref{vp} may also be formed when both resonant and nonresonant mechanisms of the formation of the nonlinear waves act simultaneously.

Eq.\eqref{vp} is the TVB oscillating with the SDFW which is the solution of the nonlinear wave equation \eqref{eq}.
This expression can be considered as a superposition of  two small amplitude breathers.
The first term of Eq.\eqref{vp} is the small amplitude ($f_{+1}$) breather oscillating with the sum of frequencies $\om+\Om_{+}$ and wave numbers $k+Q_{+}$,  and the second term is the small amplitude ($f_{-1}$) breather  oscillating with the difference of frequencies $\om-\Om_{-}$ and wave numbers $k-Q_{-}$ (taking into account Eq.\eqref{kom}). The nonlinear coupling between these two wave packets is defined by the cross terms $r_{+}|u_{-1}|^{2} u_{+ 1}$ and $r_{-}|u_{+ 1}|^{2} u_{-1}$ of Eq.\eqref{cnse}. Both breathers are polarized along the $x$ axis.
The parameters of the breathers are closely interconnected with each other and are defined by Eqs.(13), \eqref{rtw} and \eqref{rtq}.
These breathers  form the single integrated pulse Eq.\eqref{vp} which propagates in a medium with the velocity $V_{0}$  (Appendix II).
The dispersion relation and connections between the oscillating parameters $\Omega_{\pm}$ and $Q_{\pm}$ are defined by Eqs.\eqref{dis}, \eqref{diss} and (14). Similar to the scalar single-component soliton and breather, the TVB oscillating with the SDFW Eq.\eqref{vp} loses no energy while propagating through the medium.

In  view of  the fact  that in different systems the quantity $\ae_{2}$ in Eqs.(10) and (13) varies in a very wide range, we can consider in particular three different physical situations \cite{Maimistov::83,Caetano::03,Nakazawa::91,Nakazawa2::91,Adamashvili:PLA:03,Fonseca::01,Adamashvili:PhysRevE:04, Adamashvili:PRE:04, Adamashvili:PhysRevE:06,Adamashvili:Phys.Rev.A:19}.

i. In the medium where the condition  $R_{+1}<<\frac{\ae_{2}}{\Omega_{\pm }^{3}} $ is satisfied, the resonant TVB oscillating with the SDFW can be formed under the SIT condition. In that case, the TVB with the SDFW coincides with the vector $0\pi$ pulse \cite{Adamashvili:Result:11, Adamashvili:Optics and spectroscopy:2012, Adamashvili:Physica B:12, Adamashvili:Eur.Phys.J.D.:12, Adamshvili:Arxiv:2014, Adamashvili:PhysLettA:2015, Adamashvili:Optics and spectroscopy:2019, Adamshvili:Arxiv:2019, Adamashvili:Eur.Phys.J.D.:20}.

ii. The nonresonant TVB oscillating with the SDFW can be formed when the light and  medium  parameters satisfy the condition  $R_{+1}>>\frac{\ae_{2}}{\Omega_{\pm }^{3}} $ \cite{Adamshvili:Arxiv:2018}.

iii. The hybrid TVB oscillating with the SDFW can be formed in a situation where the light and  medium  parameters satisfy the requirement that the quantities $R_{+1} $  and $\frac{\ae_{2}}{\Omega_{\pm }^{3}} $ be of the same order.

Comparing  Eqs.\eqref{Ap} and \eqref{equa} we clearly see that the second-order derivatives   $g_{2} \frac{\partial^{2}\hat{E}_{l}}{\partial z^2} -g_{5} \frac{{{\pa}^2}{\hat{E}_l}}{\pa t^2} -g_{4}\frac{{{\pa}^2}{\hat{E}_l}}{{\pa {z}}{\pa  t}}$  are smaller than the first-order derivatives $i(g_{1} \frac{\partial \hat{E}_{l}}{\partial z}+g_{3} \frac{{\pa}{\hat{E}_l}}{\pa t}) $. Nevertheless the second-order derivatives  play a significant role in the process of formation of nonlinear solitary waves. The second-order derivatives of Eq.\eqref{equa} are connected with the terms $g_{2} Q_{l,n}^{2} -g_{5} {\Omega^{2}_{l,n} } +g_{4}Q_{l,n}  \Omega_{l,n}$ of Eq.\eqref{diss} and the  characteristic parameters $\Omega_{\pm}$ and $Q_{\pm}$ of Eq.(14).

In the particular case, if we neglect the second-order derivatives in Eq.\eqref{equa}, as was usually done before the development of the generalized PRM, then we can obtain  scalar single-component SIT-soliton and scalar single-component SIT-breather solutions of Eq.\eqref{equa}. Indeed, in this case, neglecting in Eq.\eqref{diss} the terms with the coefficients $g_{2}$, $g_{4}$ and $g_{5}$, we see that this equation does not any longer depend on the indexes $l$ and $n$. Consequently, the conditions $\Omega_{+}=\Omega_{-}$, $Q_{+}=Q_{-}$ and $r_{\pm}=0$ are fulfilled and the resonant TVB oscillating with the SDFW (case i)  is partitioned into two scalar single-component  SIT-breathers which propagate independently of each other.
Note, that in the theory of SIT we can neglect the second-order derivatives because they do not take part in the process of formation of SIT-pulses and take part only in the process of formation of the bound state of two breathers.

For the nonresonant case ii, the situation is different. The second-order derivatives in Eq.\eqref{equa} not only take part in the formation of the bound state of two breathers, but also take part  in the process of formation of nonresonant nonlinear solitary wave.  This is explained by the fact that for the formation of nonresonant  nonlinear solitary wave it is necessary that the Kerr-nonlinearity be balanced by  dispersion (the diffraction effects are not considered here).
Dispersion effects are connected with the term $H_{l,n}$ in Eq.(10) and if we neglect the second-order derivative terms in Eq.\eqref{equa} with the coefficients $g_{2}$, $g_{4}$, $g_{5}$ then we observe that the quantity $H_{l,n}$ becomes equal to zero and the nonresonant solitary wave is not formed. So, when we investigate nonresonant nonlinear solitary waves, in contrast to SIT theory (case i) we should  always take into consideration the second-order derivatives in wave equation \eqref{equa}.

For the hybrid TVB oscillating with the SDFW (case iii) the situation is similar to the nonresonant case because for hybrid waves both resonant and nonresonant mechanisms are involved in the formation of nonlinear solitary waves and, consequently, we  must retain  the second-order derivatives in the wave equation \eqref{equa}.

We also have to note yet another important circumstance concerning  the second-order  derivatives in the wave equation \eqref{equa}. Although  we can keep the  second-order  derivatives  in Eq.\eqref{equa}, nevertheless  we can not  always obtain  TVB  oscillating with  SDFW.  This happens so because for the investigation of two-component solitary waves it is necessary to have a sufficient number of auxiliary  complex functions  and parameters. In particular, the standard PRM \cite{Taniuti::1973}, contains  only one complex auxiliary  function  and two constant parameters which is not enough for the description of two-component solitary  waves (see, for instance \cite{Adamashvili:PRE:04,Adamashvili:PhysRevE:04,Adamashvili:Eur.Phys.J.D.:20}).  However the application of the generalized PRM Eq.\eqref{ser} gives us a chance to introduce  two complex auxiliary  functions  and eight constant parameters.

To summarize the results discussed above, we make the following  conclusions:  for the investigation of resonant, nonresonant and hybrid TVB oscillating with SDFW it is necessary to take into consideration the second-order  derivative  terms of the  wave equation \eqref{equa}, the waves must  be the  phase-modulated (see Section  III), and  the  generalized PRM Eq.\eqref{ser} should be used.

The hybrid TVB oscillating with SDFW Eq.\eqref{vp} adds to new physical conditions in which allow us to study this two-component solitary  wave  previously had been studied only for the resonance  and  nonresonanse  waves. On the other hand, together  with  hybrid  one-component solitons and  hybrid  one-component breathers, the  hybrid  TVB  oscillating with SDFW  makes the theory  of hybrid  nonlinear  solitary  waves more complete.

\vskip+0.5cm
\section{Appendix I}

\begin{equation}\label{D1}
D_{l}=\int \ve(z_1,t_1)E(z-z_1,t-t_1)dz_1 dt_1,
\end{equation}
Substituting Eq.\eqref{E} into Eq.\eqref{D1}, and taking Eq.\eqref{Ap} into consideration,  we obtain
\begin{equation}\label{D2}
D_{l}=\sum_{l}{Z_l}[\kappa -ia \frac{{\pa}}{\pa {z}}+ib
\frac{{\pa}}{\pa t} -c \frac{{{\pa}^2}}{\pa {z}^2} -d \frac{{{\pa}^2}}{\pa t^2}+ \tilde{t}\frac{{{\pa}^2}}{{\pa {z}}{\pa t }}]\hat{E}_l,
\end{equation}
where
$$
{{\kappa}}=\int
\ve(z_1,t_1)e^{-il(kz_1-\om t_1)} dz_1 dt_1,
$$
$$
a=-i\int \ve(z_1,t_1){z_1}e^{-il(kz_1-\om t_1)}
dz_1 dt_1,
$$
$$
b=i\int \ve(z_1,t_1){t_1}e^{-il(kz_1-\om t_1)}
dz_1 dt_1,
$$
$$
c=-\int
\ve(z_1,t_1)\frac{{z_1}^2}{2}e^{-il(kz_1-\om t_1)} dz_1 dt_1,
$$
$$
d=-\int
\ve(z_1,t_1)\frac{{t_1}^2}{2}e^{-il(kz_1-\om t_1)} dz_1 dt_1,
$$
$$
\tilde{t}=\int \ve(z_1,t_1){t_1}{z_1}e^{-il(kz_1-\om
t_1)} dz_1 dt_1.
$$

Substituting Eqs.  \eqref{E}  into the Bloch equations \eqref{bl}, and the using the condition of the inhomoheneouse broadening of the spectral line, we obtain a polarization of two-level atoms in the form
\begin{equation}\label{p}
P_{r}= i\frac{n_0 \mu^{2}}{\hbar} <g>
\sum_{l=\pm 1}l Z_l [\ve^{1}  {{\Theta}_{l}}^{(1)}
 +\ve^{2} {{\Theta}_{l}}^{(2)}+\ve^{3}  {{\Theta}_{l}}^{(3)}
 -\ve^{3}\frac{{{\kappa}^{2}_0}}{2}
\int\frac{\pa {{\Theta}_{l}}^{(1)}}{\pa t}{{\Theta}_{-l}}^{(1)}
{{\Theta}_{l}}^{(1)}dt']+{\cal O}(\epsilon^{4})
\end{equation}
where
\begin{equation}\label{yy}\nonumber\\
 \Theta_{l}^ {(\alpha)} (z,t)=\int_{-\infty}^t\hat{E_{l}}^ {(\alpha)}(z,t')dt',
 \;\;\;\;\;\;\;\;\;\;\;\;\;\;\;\;\;\;\;<g>= \int \frac{g(\Delta)d\Delta}{1+T^2 \Delta^2},
\end{equation}
$g(\Delta)$ is the inhomogeneous broadening lineshape function for an ensemble of two-level optical atoms or SQDs, $\Delta=\omega_{0}-\omega$.

\vskip+0.5cm

\section{Appendix II}

The analytical solution of Eq.\eqref{cnse} in the form of  two components of the vector soliton $u_{+}$ and $u_{-}$ has the form
(see, for instance \cite{Adamashvili:Optics and spectroscopy:2012, Adamashvili:Eur.Phys.J.D.:12} and references therein)
\begin{equation}\label{uet}
u_{\pm}(z,t)=\frac{f_{\pm 1}}{\mathfrak{b} T}\sech(\frac{t-\frac{z}{V_{0}}}{T}) e^{i(k_{\pm 1}z - \omega_{\pm 1} t )},
\end{equation}
where the quantities $f_{\pm 1},\; k_{\pm 1}$ and $\omega_{\pm 1}$ are the real constants.
$V_{0}$  is the  nonlinear wave velocity.
\begin{equation}\label{rtw}
T^{-2}=V_{0}^{2}\frac{v_{+}k_{+1}+k_{+1}^{2}p_{+}-\omega_{+1}}{p_{+}},\;\;\;
{\mathfrak{b}}^{2}=V_{0}^{2} \frac{f_{+1}^{2}\mathfrak{q}_{+}+f_{-1}^{2}r_{+}}{2p_{+}},\;\;\;\;\;k_{\pm 1}=\frac{V_{0}-v_{\pm}}{2p_{\pm}}.
\end{equation}
The following inequalities
\begin{equation}\label{kom}
k_{\pm 1}<<Q_{\pm },\;\;\;\;\;\;\omega_{\pm 1}<<\Omega_{\pm },
\end{equation}
are valid.

The relations between the quantities $f_{\pm 1}$ and $\omega_{\pm 1}$ have the form
\begin{equation}\label{rtq}
f_{+1}^{2}=\frac{p_{+}\mathfrak{q}_{-}- p_{-}r_{+}}{p_{-}\mathfrak{q}_{+}-p_{+}r_{-}}f_{-1}^{2},
\;\;\;\;\;\;\;\;\;
\omega_{+1}=\frac{p_{+}}{p_{-}}\omega_{-1}+\frac{V^{2}_{0}(p_{-}^{2}-p_{+}^{2})+v_{-}^{2}p_{+}^{2}-v_{+}^{2}p_{-}^{2}
}{4p_{+}p_{-}^{2}}.
\end{equation}

\vskip+0.5cm

\end{document}